\documentstyle[aps,epsfig,prb]{revtex}  			
\begin{document} 
\twocolumn[\hsize\textwidth\columnwidth\hsize\csname  	 
@twocolumnfalse\endcsname				
\draft
\title{Fermi-Liquid Theory for Anisotropic Superconductors}
\author{M. B. Walker}
\address{Department of Physics, 
University of Toronto,
Toronto, Ont. M5S 1A7 }
\date{\today }
\maketitle

\widetext					%
\begin{abstract}
This article develops a Fermi-liquid theory for superconductors with anisotropic Fermi surfaces, Fermi-liquid interactions, and energy gaps. For $d$-wave superconductors, the Fermi-liquid interaction effects are found to be 
classifiable into strong and negligible renormalizaton
effects, for symmetric and antisymmetric combinations of the 
energies of $k\uparrow$ and $-k\downarrow$ quasiparticles, respectively.  
Furthermore, the leading clean-limit 
temperature-dependent correction to the superfluid density in a $d$-wave superconductor is found to be renormalized by a Fermi velocity (or mass) renormalization effect. The question is raised of whether or not the penetration depth in the high temperature superconductor YBa$_2$Cu$_3$O$_{6+x}a$ can be accounted for with physically acceptable parameters within the framework of a quasiparticle model.  Fermi-liquid corrections to the spin susceptibility and to the zero-energy magnetic-field-induced density of states are also evaluated.
\end{abstract}


\vfill		
\narrowtext			%

\vskip2pc]	

\section{Introduction}
There is now considerable experimental evidence
that the  cuprate high $T_c$ superconductors exhibit the 
simple power law temperature dependences predicted by the
quasiparticle picture for their thermodynamic 
and transport properties
at temperatures well below $T_c$.
For example,
penetration depth measurements find that the superfluid
density exhibits a low-temperature clean-limit 
linear-in-T temperature dependence \protect\cite{har93},
in agreement with theory \protect\cite{pro91}. The NMR
relaxation rate exhibits the expected $T^3$ temperature
dependence \protect\cite{mar93}.  The predicted effect of
impurities in giving rise to a universal thermal conductivity
\protect\cite{lee93,gra96} has been confirmed \protect\cite{tai97}.
The clean-limit specific heat varying as $T^2$ appears to have
been observed \protect\cite{mol97,wri99}.
Even the electrical transport relaxation rate observed in
microwave conductivity experiments \protect\cite{hos99}, which
had resisted explanation for some time, has  now been explained
in terms of a quasiparticle picture \protect\cite{wal00}.

Whether or not the Fermi-liquid parameterization of the coefficients of the above power law temperature dependences is quantitatively accurate is at
present an open question.  A recent study correlating these
different coefficients \protect\cite{chi00} concludes that the 
quasiparticle model may be successful here also
provided a Fermi-liquid interaction
factor multiplying the superfluid density is treated as an
adjustable parameter. Some remarks at the end of this article address the question of whether or not the experimentally determined value of this adjustable parameter has a physically reasonable value.  The answer to this question should help to assess the validity of the quasiparticle picture of the low temperature properties of high T$_c$ superconductors. Recent debate on correctness of the quasiparticle picture
is also occurring in connection with ARPES experiments
\protect\cite{kam00,val99}, and
in connection with the role of phase fluctuations of the
complex order parameter 
in the determination of the temperature
dependence of the superfluid density \protect\cite{car99}.

A principal goal of this article is to develop a Fermi-liquid theory applicable to superconductors with anisotropic Fermi surfaces, Fermi-liquid interactions, and energy gaps, and to apply it to the $d$-wave superconductors at temperatures well below the critical temperature $T_c$. This theory will then be used to develop formulae, including Fermi-liquid corrections, for the London penetration depth, the spin susceptibility and the magnetic field contribution to the density of states and specific heat (observable in the mixed state).  The approach used here is a relatively elementary phenomenological one, similar in spirit to that of Landau's original article \protect\cite{lan57} and its extension to  superconductors by Betbeder-Matibet and Nozi\`{e}res \protect\cite{bet69}, and it is hoped that it will be of pedagogical interest.  An important aspect of the extension of the Landau theory of Fermi liquids to superconductors is the introduction of an appropriate dependence of the energy functional on the superfluid momentum. The classic articles of Larkin \protect\cite{lar63} and of Leggett \protect\cite{leg65} used a more formal correlation function approach than is used here. Other studies of Fermi-liquid
interactions in unconventional superconductivity include Refs.\ 
\protect\onlinecite{gro86,xu95}.

It should be emphasized that Fermi-liquid theory is known not to give an adequate description of the normal-state properties of the high $T_c$ superconductors.  Thus, in the application of the results of this article to high-temperature superconductors, it is only the properties of the superconducting state at temperatures well below the critical temperature $T_c$
that will be considered as  being possibly explicable in terms of Fermi-liquid theory.

The potential importance of Fermi-liquid interactions in 
renormalizing the superfluid density in the case of $d$-wave superconductors has been emphasized in Refs.\ \protect\onlinecite{mil98} and  
\protect\onlinecite{dur00}. The calculations of Fermi liquid properties in these articles were carried out within the framework of models with isotropic Fermi surfaces and Fermi-liquid interactions, but with $d$-wave energy gaps, and they  demonstrated the existence of a factor renormalizing the superfluid density that was given in terms of the isotropic Fermi-liquid interaction parameter $F_1^s$.  
In a preliminary account of the present work
\protect\cite{wal00b} a similar factor renormalizing the superfluid density was found, but expressed as a ratio of two different types of velocities (i.e. roughly speaking as an effective mass ratio).  By using the Landau effective mass relation \protect\cite{lan57} known to be valid for Galilean invariant systems, these two results could be seen to be equivalent, but the relation between the two results for the case of systems with an underlying crystal lattice was far from clear, and has been the subject of some discussion and controversy.  Similarly, the expression used to calculate the current (analogous to Eq.\ \ref{Jsup} below) in Ref.\ \protect\onlinecite{wal00b} has 
been controversial.  For these reasons, the basic ideas of the approach to the Fermi-liquid theory of anisotropic superconductors used in this article are outlined in some detail in Sections II and III below. The formula developed below for the penetration depth is shown the be in agreement with the results of Larkin \protect\cite{lar63} and of Leggett \protect\cite{leg65} for $s$-wave superconductors with isotropic Fermi surfaces and Fermi-liquid interactions, and with those of Refs.\ \protect\onlinecite{mil98} and  
\protect\onlinecite{dur00} for $d$-wave superconductors with isotropic Fermi surfaces and Fermi-liquid interactions.  The result below for the low-temperature penetration depth for $d$-wave superconductors is also valid for the case of anisotropic Fermi surfaces and Fermi-liquid interactions. A formula applicable to this case has also been developed independently and stated in Ref.\ \protect\onlinecite{ran00}, which finds a result expressed in terms of parameters that are different from those in our formula; there is thus the possibility that these two results are not equivalent, as a result of the absence of a generalization of the Landau effective mass relation to the case of
an anisotropic Fermi surface. 

The renormalization of generalized external fields by Fermi-liquid interactions in a $d$-wave superconductor at low temperatures has an interesting symmetry property. This manifests itself when the quasiparticle energies
are separated into parts that are ${\mathcal{S}}$ymmetric and 
${\mathcal{A}}$ntisymmetric combinations of the energies of the
$+k\uparrow$ and $-k\downarrow$ states. (The calligraphic letters
${\mathcal{S}}$ and ${\mathcal{A}}$ are used here to emphasize the
difference with the more usual definition of the symmetric and 
antisymmetric combinations with respect to $+k\uparrow$ and
$+k\downarrow$ states common in normal state analyses,
e.g. see Eq. 1.32 of Ref.\ \protect\onlinecite{pin89}.) 
In the presence of
Fermi-liquid interactions, the ${\mathcal{S}}$ymmetric and 
${\mathcal{A}}$ntisymmetric corrections to the quasiparticle
energies obey integral equations that are independent of each other,
and they are renormalized differently.  This leads to the fact that the 
${\mathcal{S}}$ymmetric 
external fields exhibit strong Fermi-liquid renormalization effects to which quasiparticles from the entire Fermi surface contribute (cf. Ref.\ \protect\onlinecite{mil98}), 
while the the ${\mathcal{A}}$ntisymmetric external fields 
exhibit relatively weak temperature-dependent renormalizations that arise from the nodal quasiparticles only, and that
can often be neglected.

Temperature gives a ${\mathcal{S}}$ymmetric correction to the
quasiparticle energy because $+k\uparrow$ and $-k\downarrow$ states
are affected in the same way by temperature.  A superfluid flow
generates an ${\mathcal{A}}$ntisymmetric correction since the
components of $+k$ and $-k$ along the superfluid velocity have
opposite signs.  Also the Zeeman interaction generates an 
${\mathcal{A}}$ntisymmetric correction because the spin $\uparrow$
and spin $\downarrow$ contributions to the energy have opposite 
signs.  Thus the superfluid density and the magnetic susceptibility
are negligibly renormalized by Fermi-liquid interactions, while
the effects of temperature (although relatively small) are strongly
renormalized by Fermi-liquid interactions.

\section{Hamiltonian and Current Density}

The Hamiltonian describing a system of electrons in a periodic lattice potential
$V_{per}({\bf r})$, a magnetic field described in terms of a vector potential ${\bf A(r)}$, and interaction through an electron-electron interaction ${\mathcal  H}_{int}$ is
\begin{equation}
	{\mathcal H} = \int d^3 r \Psi_\sigma ^\dagger ({\bf r}) H_0 
	\Psi_\sigma({\bf r}) + {\mathcal H}_{int}
	\label{Ham}
\end{equation}
where
\begin{equation}
	H_0 = \frac{1}{2m}\left ( -i\hbar \nabla + {\bf p}_s \right )
		+ V_{per}({\bf r}) - \mu
	\label{H_0}
\end{equation}
and
\begin{equation}
	{\bf p}_s =  \hbar \nabla (\theta/2) - \frac{e}{c} {\bf A}.
	\label{p_s}
\end{equation}
Note that a term $-\mu N$ has been included in $\mathcal{H}$.
Also, the current density operator is given by
\begin{equation}
	\hat{\bf J}({\bf r}) = \frac{e}{2m} 
	\left\{ 
	\left[(-i\hbar \nabla + \bf{p}_s)\Psi_\sigma ({\bf r}) \right ] ^\dagger
	\Psi_\sigma ({\bf r}) + {\rm h.c.} 
	\right\}
	\label{cur}
\end{equation}
where h.c. indicates the Hermitian conjugate of the preceding term.
The above four equations are probably more commonly written as above but
with $\theta = 0$.  The substitution $\Psi_\sigma ({\bf r}) \rightarrow
\Psi_\sigma ({\bf r})e^{i\theta/2}$ in the more common form of these 
equations yields the equations as written.  In the superconducting state,
${\bf p}_s$, which is a gauge-invariant quantity, will be interpreted as the superfluid momentum. In calculations of the penetration depth in the London limit below, ${\bf p}_s$ can be assumed to be uniform.

The thermodynamic potential $\Omega$ and the equilibrium density operator 
$\rho$ are given by 
\begin{equation}
	\Omega = - k_B T {\rm ln} \left( {\rm tr} e^{-\beta \mathcal{H}}
		\right),\ \ 
	\rho = \frac{e^{-\beta \mathcal{H}}}{{\rm tr} e^{-\beta \mathcal{H}}}.
	\label{Omega}
\end{equation}
It is easily seen that the average value of the current density operator,
defined by ${\bf J(r)} = {\rm tr} \rho {\bf \hat{J}(r)}$, can 
be calculated by taking the functional derivative of the thermodynamic
potential $\Omega$ with respect to the vector potential $\bf A(r)$, i.e.
\begin{equation}
	{\bf J(r)} = -c \delta \Omega / \delta {\bf A(r)}.
	\label{JbyA}
\end{equation}
If the current density is homogeneous, which will be the case of direct 
interest below, it can be calculated by using the formula
\begin{equation}
	{\bf J} = \frac{1}{V} \int d^3 r {\rm tr} \rho {\bf \hat{J}(r)}
		=\frac{e}{V} \frac{\partial \Omega}{\partial {\bf p}_s}.
	\label{Jbyp}
\end{equation}

The energy eigenvalues of the Hamiltonian $H_0$ for
${\bf p}_s = 0$, which describes a single electron in a periodic lattice
potential, form energy bands with the allow states 
characterized by a wave vector ${\bf k}$ lying in the
first Brillouin zone.  Subsequent discussion will consider only a single
energy band, whose energy spectrum is described by the function 
$\varepsilon^b(\hbar{\bf k})$. The general problem of the energy spectrum of
an electron in an external magnetic field is quite complex, but for sufficiently weak magnetic fields, i.e. a sufficiently slowly varying vector potential, a quasiclassical approximation can be developed which gives the energy spectrum of a single band of electrons in the form $\varepsilon^b_k ({\bf p}_s)$ =
$\varepsilon^b(\hbar {\bf k} + {\bf p}_s)$\protect\cite{lif}. The single-band Hamiltonian envisaged in this article thus has the
form
\begin{eqnarray}
	{\mathcal H}^b & = & \sum_{k \sigma}  
	\varepsilon^b_k({\bf p}_s) c^\dagger_{k \sigma} c_{k \sigma} +
		\nonumber \\
	& & \frac{1}{2V} \sum_{k_1 k_2 k_3 k_4 \sigma_1 \sigma_2}
	V_{k_1 k_2 k_3 k_4}c^\dagger_{k1 \sigma_1} c^\dagger_{k2 \sigma_2}
	c_{k3 \sigma_2} c_{k4 \sigma_1}.
	\label{Hband}
\end{eqnarray}
The interaction term has been constructed so as to be isotropic in spin space (i.e. no spin-orbital interactions are considered), and momentum conservation to within a reciprocal lattice vector is implicitly assumed.

\section{Fermi-liquid liquid theory for anisotropic superconductors}

As an introduction to Fermi-liquid theory for superconductors, a Hartree-Fock-BCS approximation is developed.  This gives an approximation for the
parameters occurring in the Fermi-liquid theory which has some interest, and
also serves to give an indication of the structure expected for a full Fermi-liquid model.  The Hartree-Fock-BCS approximation follows from averaging the Hamiltonian of Eq.\ \ref{Hband} over a BCS type of state, i.e. by assuming a
state in which the excitations are independent Bogoliubov quasiparticles.  This gives, for the average energy (minus $\mu N$)
\begin{eqnarray} 
	E = \langle H^b \rangle & = & \sum_{k\sigma} \varepsilon^b_k({\bf p}_s) 
		n_{k\sigma}
		+ \frac{1}{2V} \sum_{k\sigma k^\prime \sigma^\prime}
		f_{k k^\prime}^{\sigma \sigma^\prime}
		n_{k\sigma}n_{k^\prime \sigma^\prime} +
		\nonumber	\\
		&  & \frac{1}{V} \sum_{k k^\prime} g_{k k^\prime} W_k^\ast 				W_{k^\prime}
	\label{HFBCS}
\end{eqnarray}
where
\begin{equation}
	f_{k k^\prime}^{\sigma \sigma^\prime} = V_{k k^\prime k^\prime k}
	- \delta_{\sigma \sigma^\prime} V_{k k^\prime k k^\prime}, \ \ 
	g_{k k^\prime} = V_{k \bar{k} k^\prime \bar{k^\prime}}
	\label{fg}
\end{equation}
with $\bar{k} = -k$ and
\begin{equation}
	n_{k \sigma} = \langle 	c^\dagger_{k \sigma} c_{k \sigma} \rangle, \ \ 
	W_k = \langle c_{k \uparrow} c_{-k \downarrow} \rangle.
	\label{nW}
\end{equation}

In Fermi-liquid theory, the energy is written in terms of the ground state energy, plus a contribution due to excitations from the ground state.  Thus,
in Eq.\ \ref{HFBCS} put $n_{k\sigma} = n^0_{k} + \delta n_{k\sigma}$,
where $n^0_{k}$ is the ground state value of $n_{k\sigma}$ in the superconductor. This gives the average energy in the form
\begin{eqnarray}
	E & =  & E_0 ({\bf p}_s) +  \sum_{k\sigma} 
	\varepsilon_{k\sigma}({\bf p}_s) 
		\delta n_{k\sigma} + \nonumber \\
	 & & \frac{1}{2V} \sum_{k\sigma k^\prime \sigma^\prime}
	f_{k k^\prime}^{\sigma \sigma^\prime}
	\delta n_{k\sigma}  \delta n_{k^\prime \sigma^\prime} +
	\frac{1}{V} \sum_{k k^\prime} g_{k k^\prime} W_k^\ast W_{k^\prime}.
	\label{EFL}
\end{eqnarray}
Here
\begin{equation}
	\varepsilon_{k}({\bf p}_s) 
	= \varepsilon_{k} + {\bf v^t_k\cdot p}_s.
	\label{eps}
\end{equation}

In the Hartree-Fock-BCS approximation the energy of a quasiparticle $\varepsilon_k$ in the absence of other excited quasiparticles, and what is called here the electrical transport velocity ${\bf v}^t_k$, are given by 
\begin{equation}
	\varepsilon_{k} = \varepsilon^b_k + \frac{1}{V} \sum_{k^\prime 	\sigma^\prime}
	f_{k k^\prime}^{\sigma \sigma^\prime} n^0_{k^\prime}, \ \
		{\bf v}^t_k = \frac{1}{\hbar} 
	\frac{\partial \varepsilon^b_k}{\partial {\bf k}}.
	\label{ek}
\end{equation}
Notice from Eq.\ \ref{ek}  that in the Hartree-Fock-BCS approximation the quasiparticle energy, and hence the quasiparticle velocity 
${\bf v}_k = \hbar^{-1}\partial \varepsilon_k/\partial {\bf k}$, are changed from the bare band energy and band velocity by a term due to the electron-electron interaction, but that the transport velocity is unchanged by the electron-electron interaction.  

Consider a band kinetic energy of the form $\varepsilon^b_k = p^2/2m_b$ (in this case the effect of the periodic potential is represented by replacing the free electron mass $m$ by a band mass $m_b$) so that the basic one-electron states are the plane waves $exp(i{\bf k \cdot r})$.  Also assume an electron-electron interaction $v({\bf r}_1 - {\bf r}_2)$.  Then the Hartree-Fock
approximation to $f_{k k^\prime}^{\sigma \sigma^\prime}$ will have the form $f^{\sigma \sigma^\prime}({\bf k - k^\prime})$.  In this case the equation for ${\bf v}_k$ obtained from the first of Eqs.\ \ref{ek} can be transformed into the relation
\begin{equation}
	{\bf v}_k = {\bf v}_k^b - \frac{1}{V} \sum_{k^\prime}
		f^{\sigma \sigma^\prime}({\bf k - k^\prime})
		\frac{\partial n^0_{k\prime}}{\partial \varepsilon_{k^\prime}} 
		{\bf v}_{k^\prime}
	\label{Lemr}
\end{equation}
The model just described has the property of Galilean invariance, although
for particles of mass $m_b$ rather than the free electron mass, and Eq.\ \ref{Lemr} is the Landau effective mass relation that follows from this property \protect\cite{lan57}. For wave vectors ${\bf k}$ on the Fermi surface, ${\bf v}^b_k = \hbar {\bf k}/m_b$, while the relation 
${\bf v}_k = \hbar {\bf k}/m^\ast$ defines the quasiparticle effective mass $m^\ast$.  In the general case of an anisotropic Fermi surface, where the quasiparticle interaction function $f_{k k^\prime}^{\sigma \sigma^\prime}$ depends separately on ${\bf k}$ and ${\bf k}^\prime$, no relation analogous to
Eq.\ \ref{Lemr} can be derived. By using
${\bf v}_k = \hbar^{-1}\partial \varepsilon_k/\partial {\bf k}$ and the Hartree-Fock result Eq.\ \ref{ek}, an integral equation is found that presumably can be solved to find ${\bf v}_k$ in terms of ${\bf v}_k^b$ and the Fermi-liquid interaction parameters if these are known at all points within the Fermi surface, but it is not clear that this would be a useful approach for a more
general Fermi-liquid theory.

For models that
are isotropic and Galilean invariant it is a known exact result (e.g. not restricted to the Hartree-Fock approximation just discussed) 
that the transport velocity is not renormalized by electron-electron
interactions, and hence that the transport velocity is $\hbar {\bf k}/m$
where m is the free electron mass when there is no periodic potential, or
the so-called crystalline mass $m_b$ in a model in which the effect of the
periodic potential is represented by replacing the free electron mass by
the crystalline mass in the kinetic energy \protect\cite{bet69}.  Thus the electrical transport velocity is quite a different quantity from the quasiparticle
velocity.	

If the electron-electron interactions are sufficiently strong, the 
Hartree-Fock-BCS approximation is not adequate.  Nevertheless, the energy $E$
of the system can be expected to have the same form as in Eqs.\ \ref{EFL} and
\ref{eps}, and with the parameters $f_{k k^\prime}^{\sigma \sigma^\prime}$,
$g_{k k^\prime}$, $\varepsilon_k$ and ${\bf v}^t_k$ subject to the constraints of symmetry but otherwise to be treated as
unknown parameters, some of which might be determined from measured experimental properties.  Also, the ground state energy, which must be an even function of
${\bf p}_s$, is parameterized in the form
\begin{equation}
	E_0({\bf p}_s) = E_0(0) + \frac{1}{2} V\sum_{\alpha \beta} S_{\alpha \beta}
		p_{s\alpha} p_{s\beta}
	\label{E0}
\end{equation}

An assumption implicit in this approach to the Fermi-liquid theory of superconductors is that the normal-state quasiparticle properties are not changed much in the superconducting state \protect\cite{bet69}, given that
the maximum gap is much smaller than the Fermi energy.   This 
assumption is fulfilled more strongly for conventional superconductors than for the high T$_c$ superconductors which will be the principal interest below.
However, in the investigation below of the properties of high superconductors at temperatures much lower that the maximum gap, the gap will be approximately constant, and this assumption should not give difficulties. 

The distribution function can be written \protect\cite{bet69} as the
two by two Hermitian matrix
\begin{equation}
	\hat{n}_k = \left[
	\begin{array}{cc}
		n_{k\uparrow} & W_k^\ast	\\
		W_k		& 1-n_{k\downarrow}
	\end{array}	\right]
	\label{n}
\end{equation}
so that the variation of the energy (Eq.\ \ref{EFL}) with respect to variations in the distribution function is
\begin{equation}
	\delta E = \sum_k tr ( \hat{\tilde{\varepsilon}}_k \delta \hat{n}_k).
	\label{deltaE}
\end{equation}
Here the matrix representing the quasiparticle energy in the presence of other quasiparticles is given by
\begin{equation}
	\hat{\tilde{\varepsilon}}_k = \left [
		\begin{array}{cc}
		\tilde{\varepsilon}_{k\uparrow}({\bf p}_s) &\Delta_k^\ast	\\
		\Delta_k	&	
		-\tilde{\varepsilon}_{k\downarrow}({\bf p}_s)
		\end{array} \right ]
	\label{ehat}
\end{equation}
where $\tilde{\varepsilon}_{k\sigma}({\bf p}_s) =
		 \varepsilon_{k\sigma}({\bf p}_s)
		+\delta \varepsilon_{k\sigma}$,
in which
\begin{equation}
 		\delta \varepsilon_{k\sigma} =
		+\frac{1}{V} \sum_{k^\prime \sigma^\prime}
		f_{k k^\prime}^{\sigma \sigma^\prime} 
		\delta n_{k^\prime \sigma^\prime};
	\label{deltae}
\end{equation}
also,
\begin{equation}
	\Delta_k = \frac{1}{V} \sum_k g_{kk^\prime} W_{k^\prime}.
	\label{Delta}
\end{equation}

The entropy of the Fermi liquid is 
\begin{equation}
	S = -k_B \sum_k tr [( \hat{n}_k ln \hat{n}_k 
		+ (1-\hat{n}_k)ln(1-\hat{n}_k)]
	\label{entropy}
\end{equation}
and the thermodynamic potential is $\Omega = E - TS$ (recall that E includes the term $-\mu N$). Minimizing the thermodynamic potential with respect to variations in the distribution function yields the equilibrium distribution function, which is
\begin{equation}
	\hat{n}_k = [exp(\hat{\tilde{\varepsilon}}_k/k_B T) + 1]^{-1}.
	\label{neq}
\end{equation}

Now consider the derivation of a general expression for the current density
in a superconductor by using Eq.\ \ref{Jbyp}.  It should be noted the the
thermodynamic potential depends on ${\bf p}_s$ implicitly through the dependence of the distribution function on ${\bf p}_s$, and explicitly through the dependence of $E_0({\bf p}_s)$ and $\varepsilon_{k}({\bf p}_s)$ in Eq.\ \ref{EFL} on ${\bf p}_s$.  Because the condition for equilibrium is that the derivative of $\Omega$ with respect to the distribution function is zero, the implicit dependence of the distribution function on ${\bf p}_s$ can be ignored in applying Eq.\ \ref{Jbyp}.  The current density from Eq.\ \ref{Jbyp} is thus
\begin{equation}
	J_\alpha = e \sum_\beta S_{\alpha \beta}p_{s\beta}
		+ \frac{e}{V} \sum_{k\sigma} v^t_{k\alpha} \delta n_{k\sigma}
	\label{Jsup}
\end{equation}
The first term is a temperature-independent contribution 
due to the flow of the condensate, and the second is the contribution of the excited quasiparticles.

\section{Renormalization of external fields -- basic equations}

When an external field is applied to a Fermi liquid, it excites quasiparticles,
and these quasiparticles renormalize the contribution of the external field
to the quasiparticle energy.  This is the effect that must be evaluated in
order to evaluate such effects as the screening of an external magnetic field by the superconductor (and hence the penetration depth), and the spin susceptibility. 

The quasiparticle Hamiltonian describing the excitations of the superconducting 
state has the following form:
\begin{equation}
{\mathcal{H}} = \sum_k
\begin{array}{cc}
	[c_{k \uparrow}^\dagger & c_{-k \downarrow}]
\end{array}
\left[ \begin{array}{cc}
		\zeta_k +\lambda_k & \Delta_k \\
		\Delta_k & -\zeta_k + \lambda_k
	\end{array}	\right]
\left[ \begin{array}{c} 
		c_{k \uparrow} \\
		 c_{-k \downarrow}^\dagger 
	\end{array}	\right].
\label{H}
\end{equation}
where the two by two matrix is the matrix $\hat{\tilde{\varepsilon}}_k$
of Eq.\ \ref{ehat} generalized to include arbitrary external fields,
and reorganized into ${\mathcal S}$ymmetric and  
${\mathcal A}$ntisymmetric components.
Here $\zeta_k = \varepsilon_k + \delta \varepsilon_k^{\mathcal{S}} + 
h_k^{\mathcal{S}}$, 
$\lambda_k = \delta \varepsilon_k^{\mathcal{A}} + 
h_k^{\mathcal{A}}$, and $\Delta_k$ is the momentum-dependent
gap function appropriate for $d$-wave symmetry, which is taken to be
real.  An important step in the analysis, as described
qualitatively above, is the separation of 
the excitation energies into ${\mathcal{S}}$ymmetric 
and ${\mathcal{A}}$ntisymmetric parts defined by
\begin{equation}
	\delta \varepsilon_k^{\mathcal{A}} = \frac{1}{2}
	\left[ 	\delta\varepsilon_{k\uparrow} - 	
	\delta\varepsilon_{-k\downarrow} \right],\ \ 
	\delta \varepsilon_k^{\mathcal{S}} 
	= \frac{1}{2}\left[ 	\delta\varepsilon_{k\uparrow} 
	+ \delta\varepsilon_{-k\downarrow} \right].
\label{saqpint}
\end{equation}

The quantities $h_k^{\mathcal{A}}$ and $h_k^{\mathcal{S}}$ in 
$\mathcal{H}$ represent
generalized external fields.  For example,  the case of an 
external magnetic field acting on the orbital motion of the electrons
corresponds to the external field 
$h_k^{\mathcal{A}} =  {\bf v}^t_k \cdot {\bf p}_s,\ h_k^{\mathcal{S}} 
= 0$ (cf. Eq.\ \ref{eps}).  The case of an  external magnetic field H acting on the 
spin degrees of freedom is described by taking $h_k^{\mathcal{A}} 
= \mu_B H,\ h_k^{\mathcal{S}} = 0$. In both of these cases, the 
magnetic field acts only on the ${\mathcal{A}}$ntisymmetric 
mode, and
has no effect on the ${\mathcal{S}}$ymmetric mode of excitation.

In addition to causing changes in the 
energy of a quasiparticle (as in Eq.\ \ref{deltae}), excited quasiparticles
can give rise to changes in the gap function \protect\cite{bet69}.
There are however no changes that are linear in the superfluid momentum, and this effect  will therefore be neglected. (This follows from Eqs.\ \ref{Hdiagonal} and \ref{gap}.  See also Ref.\ \protect\onlinecite{xu95}.)

The diagonalization of the Hamiltonian of Eq.\ \ref{H}  gives
\begin{equation}
{\mathcal{H}} =
\sum_{k\sigma} E_{k, \sigma} \gamma_{k, \sigma}^\dagger 
	\gamma_{k, \sigma},\ \ 
	E_{k, \sigma} = E_{\sigma k} 
	+ \sigma(\delta \varepsilon_{\sigma k}^{\mathcal{A}} 
	+ h_{\sigma k}^{\mathcal{A}})
\label{Hdiagonal}
\end{equation}
where $\sigma = \pm 1$, $E_k = \sqrt{\zeta_k^2 + \Delta_k^2}$,
and the $\gamma_{k, \sigma}^\dagger$ are operators creating
Bogoliubov quasiparticles.

\begin{figure}
\centerline{\epsfig{file=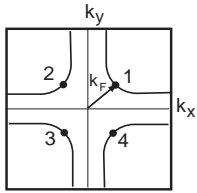,height=1in,width=1in}}
\vspace{10pt}
\caption{ The labelling of the nodes on the Fermi surface
	of YBa$_2$Cu$_3$O$_{6+x}$}
\label{fig1}
\end{figure}

Later, the energy $E_k^{(0)} = \sqrt{\xi_k^2 + \Delta_k^2}$ 
describing the quasiparticle spectrum in the absence of other 
excited quasiparticles is also used.  
For a $d$-wave superconductor,
the quasiparticle energy can be parameterized 
\protect\cite{lee93} in the neighborhood of the
Fermi-surface nodal points (see Fig.\ \ref{fig1})
as $E_k^{(0)} 
= \sqrt{(p_1 v_F)^2 + (p_2 v_2)^2}$, where $p_1$ and 
$p_2$ are components of the momentum relative to the 
nodal point in directions perpendicular and parallel 
to the Fermi line.
At low temperatures, only quasiparticles close to these 
four points can be thermally excited. 

Using Eq.\ \ref{deltae} in Eq.\ \ref{saqpint} and keeping 
only terms up to linear order in the $\delta \varepsilon$'s 
and $h$'s yields 
the integral equations
\begin{equation}
	\delta \varepsilon_k^{\mathcal{S}}  =  \frac{2}{V}
	\sum_{k^\prime} f^{(+)}_{k k^\prime} 	
	    \left[ \frac{\varepsilon_{k^\prime}}{E_{k^\prime}^{(0)}}
	f(E_{k^\prime}^{(0)}) 
	-\frac{\Delta_{k^\prime}^2}{E_{k^\prime}^{(0)3}} 
	(\delta \varepsilon_{k^\prime}^{\mathcal{S}} 
	+ h_{k^\prime}^{\mathcal{S}}) 	\right]
	\label{S}
\end{equation}
and
\begin{equation}
	\delta \varepsilon_k^{\mathcal{A}} = \frac{2}{V}\sum_{k^\prime} 
	f^{(-)}_{k k^\prime} \frac{\partial f}{\partial E_{k^\prime}^{(0)}}
	\left(\delta \varepsilon_{k^\prime}^{\mathcal{A}} 
	+ 	h_{k^\prime}^{\mathcal{A}} \right),	
	\label{A}
\end{equation}	
where interaction parameters $f^{(\pm)}_{k k^\prime}$ (in contrast to the more conventionallly defined\protect\cite{pin89} parameters $f^{s,a}_{k k^\prime}$) are defined by
\begin{equation}
	f^{(\pm)}_{k k^\prime} = \frac{1}{2}(f^{\sigma \sigma}_{k k^\prime}\pm
		f^{\sigma \overline{\sigma}}_{k,-k^\prime}),\ \
	f^{s,a}_{k k^\prime} = \frac{1}{2} (f^{\sigma \sigma}_{k k^\prime} 
		\pm f^{\sigma \overline{\sigma}}_{k k^\prime}).
	\label{fsa}
\end{equation}
In applying these equations to a copper-oxide plane of a high T$_c$ superconductor, the volume $V$ should be replaced by the area $L^2$. Also,
$f(\varepsilon)=[exp(\varepsilon/k_BT)+1]^{-1}$.

The gap equation can be found from Eqs.\ \ref{nW}, \ref{Delta}
and \ref{Hdiagonal}, to be
\begin{equation}
	\Delta_k = \frac{1}{2V} \sum_{k^\prime} g_{kk^\prime}
	\frac{\Delta_{k^\prime}}{E_{k^\prime}}
	[1-f(E_{k^\prime 1}) - f(E_{-k^\prime,-1})].
	\label{gap}
\end{equation}

\section{Penetration depth for isotropic $s$-wave superconductors}
The formalism just developed can be tested by deriving a formula for the penetration depth in the London limit for the case of a not necessarily translationally invariant model having an isotropic Fermi surface, Fermi liquid interactions and energy gap, and comparing the results with those obtained in the classic articles by Larkin\protect\cite{lar63} and by Leggett\protect\cite{leg65}. To do this, consider Eq.\ \ref{A} with the external field $h_k^{\mathcal{A}} =  {\bf v}^t_k \cdot {\bf p}_s$ describing the interaction of the orbital motion of the quasiparticle with a magnetic field.  Because $h_k^{\mathcal{A}}$ is odd in {\bf k}, the parameter $f^{(\pm)}_{k k^\prime}$ can be transformed into
$f^s_{k k^\prime}$.  Also, for an isotropic Fermi surface
\begin{equation}
	\varepsilon_k = \hbar(k-k_F)v_F,\ \
	 f^{s,a}_{k k^\prime} = \sum_{\ell=0}^{\infty} f^{s,a}_\ell P_{\ell}
		[cos({\bf k\cdot k^\prime}/k_F^2)].
	\label{fsal}
\end{equation}
The solution to the integral equation \ref{A} now gives the renormalized value of the external field $h^{\mathcal A}_k = {\bf v}^t_k \cdot {\bf p}_s$ as
\begin{equation}
	h^{\mathcal A}_k + \delta \varepsilon^{\mathcal A}_k   = 
	 {\bf v}^t_k \cdot {\bf p}_s/
	[1 + \frac{1}{3}F_1^sf(T)]
	\label{deltaeiso}
\end{equation}
where
\begin{equation}
	F_1^s = 2\nu(0) f_1^s ,\ \
	f(T) = -\frac{1}{\nu (0) V}\sum_k 
		\frac{\partial f}{\partial E_k^{(0)}}, 	
	\label{defF}
\end{equation}
and $\nu(0) = k_F^2/(2\pi^2\hbar v_F)$.  The quantity $f(T)$ can be regarded\protect\cite{leg65} as an ``effective density of single-particle levels'' near the Fermi surface relative to the normal-state value $\nu(0)$.  Note that $f(T_c) = 1$ and $f(0) = 0$. The contributions
$\delta \varepsilon_k^{\mathcal{S}}$ to the quasiparticle energy will be assumed
to be negligible because of the presence of the factor $\varepsilon_k$ on the right hand side of Eq.\ \ref{S}.

A more restricted isotropic model is now considered in which, following Betbeder-Matibet and Nozi\`{e}res\protect\cite{bet69}, the band energy is assumed to have the form $\varepsilon_k^b = \hbar^2 k^2/(2m_b)$, i.e. the effect of the underlying periodic potential is simulated by introducing a crystalline mass $m_b$ in the place of the free electron mass.  The important thing here is that the kinetic energy is quadratic in momentum, which is a necessary condition for Galilean invariance.  Secondly, it should be noted that the assumption of isotropy (rotational invariance with respect to arbitrarily placed rotation axes) automatically implies invariance with respect to an arbitrary translation, since a product of two rotations about different but parallel axes  through equal but opposite angles is equivalent to a translation.  Thus isotropy  implies translational invariance, and these symmetries, coupled with a kinetic energy quadratic in momentum imply Galilean invariance.  Hence the Landau effective mass relation \protect\cite{lan57} corresponding to non-interacting particles of mass $m_b$ applies, which yields $v^t_F = v_F(1 + F_1^s/3)$.  This
relation will be used to eliminate $v^t_F$ from future formulae.

Now using Eq.\ \ref{deltaeiso} together with 
Eq.\ \ref{Jsup} yields, for the inverse square London penetration depth,
\begin{equation}
	\frac{1}{\lambda_L^2} = 
		\frac{4e^2 k_F^2 v_F}{3\pi \hbar c^2}
		(1+\frac{F_1^s}{3})
	\left[ 1 - (1 + \frac{F_1^s}{3})  \frac{f(T)}{1+\frac{1}{3}F^s_1 f(T)}
	\right].
	\label{lambdaiso}
\end{equation}
It is the second term in
Eq.\ \ref{Jsup} (i.e. the quasiparticle contribution to the current) that gives the second term in the square brackets in the above equation (which is the temperature dependent contribution to the inverse square penetration depth). The
temperature independent first term in the square brackets in the above equation,
i.e. unity, should contain a factor $S_{\alpha \beta}$ (from the first term in
Eq.\ \ref{Jsup}).  This factor 
$S_{\alpha \beta}$ has been eliminated by imposing the condition that the inverse square penetration depth should go to zero at $T = T_c$ where $f(T_c) = 1$.

Eq.\ \ref{lambdaiso} for the penetration depth can be seen to be identical to the  result of Larkin\protect\cite{lar63} at $T=0$, and to Eq. 70 of Leggett
\protect\cite{leg65} at general $T$ in the superconducting state.  The model
analyzed in this section is perhaps not, however, identical to the models used
in those articles.  For example Leggett states that he considers a model that is
isotropic, but at the same time states that translational invariance is not
assumed while recognizing that this can only be an approximation.  Thus, the basic assumptions of our work and those of Refs.\ \protect\onlinecite{lar63} and
\protect\onlinecite{leg65}, while related, may be somewhat different.  Fortunately the end results agree.

\section{Application to $d$-wave Superconductors}
Now consider a $d$-wave high-temperature superconductor modeled by a single
copper-oxide plane having a Fermi surface such as that shown in Fig.\ \ref{fig1}.  From Eq.\ \ref{A}, which determines the 
${\mathcal{A}}$ntisymmetric corrections to the 
quasiparticle energies, it is clear that only the 
values of $\delta \varepsilon_k^{\mathcal{A}}$ and 
$h_k^{\mathcal{A}}$ at the Fermi surface nodes are 
relevant to the low energy properties (i.e. when $k_B T$ is much less than
the maximum gap).  Also, the 
solutions of Eq.\ \ref{A} can be classified according 
to the irreducible representation of the point group 
$C_{4v}$ (or 4mm) describing a tetragonal copper-oxide 
plane of a high T$_c$ superconductor, the independent 
solutions being
\begin{eqnarray}
	\delta \varepsilon^{\mathcal{A}}_{A_g} &=& 
	\left( \delta \varepsilon^{\mathcal{A}}_1 +
	\delta \varepsilon^{\mathcal{A}}_2 
	+ \delta \varepsilon^{\mathcal{A}}_3 
	+ \delta \varepsilon^{\mathcal{A}}_4 \right) /4 \nonumber \\
	\delta \varepsilon^{\mathcal{A}}_{xy} &=& 
	\left( \delta \varepsilon^{\mathcal{A}}_1 
	- \delta \varepsilon^{\mathcal{A}}_2 
	+ \delta \varepsilon^{\mathcal{A}}_3 
	- \delta \varepsilon^{\mathcal{A}}_4 \right) /4 \nonumber \\
	\delta \varepsilon^{\mathcal{A}}_{Ex} &=& 
	\left( \delta \varepsilon^{\mathcal{A}}_1 
	- \delta \varepsilon^{\mathcal{A}}_2 
	- \delta \varepsilon^{\mathcal{A}}_3 
	+ \delta \varepsilon^{\mathcal{A}}_4 \right) /4 \nonumber \\
	\delta \varepsilon^{\mathcal{A}}_{Ey} &=& 
	\left( \delta \varepsilon^{\mathcal{A}}_1
	+ \delta \varepsilon^{\mathcal{A}}_2 
	- \delta \varepsilon^{\mathcal{A}}_3 
	- \delta \varepsilon^{\mathcal{A}}_4 \right)/4 
	\label{irreps}
\end{eqnarray}
where the indices 1,2,3 and 4 refer to the four nodes in the 
excitation spectrum, as defined in Fig.\ \ref{fig1}.
The external 
fields $h^{\mathcal{A}}_k$ at the nodes can be similarly classified.

The solution of Eq.\ \ref{A} now gives the renormalized
$h_{\Gamma}^{\mathcal{A}}$ as
\begin{equation}
	h_{\Gamma}^{\mathcal{A}} + \delta \varepsilon_{\Gamma}^{\mathcal{A}}(T)  
	= h_{\Gamma}^{\mathcal{A}}/[1 + F^{\mathcal{A}}_{\Gamma}(T)]
	\label{delta_e}
\end{equation}
with $F^{\mathcal{A}}_{\Gamma}(T) = f^{\mathcal{A}}_{\Gamma} 
ln(2) k_B T/(2\pi \hbar^2 v_F v_2)$.  For $d$-wave superconductors, the quantity $v_F$ is the magnitude of the quasiparticle velocity at a node, as discussed following Eq.\ \ref{Hdiagonal}. 
Also, $\Gamma$ represents any of the irreducible representations present
in Eqs.\ \ref{irreps}.  The $f^{\mathcal{A}}_{\Gamma}$'s are defined by
\begin{eqnarray}
	f^{\mathcal{A}}_{A_g} 
		&=& f^a_{11} + f^a_{13} + 2 f^a_{12} \nonumber \\
	f^{\mathcal{A}}_{xy} 
		&=& f^a_{11} + f^a_{13} - 2 f^a_{12} \nonumber \\
	f^{\mathcal{A}}_{E} &=& f^s_{11} - f^s_{13}.
	\label{fGamma}
\end{eqnarray}
where $f^a_{12}$ for example is
$f^a_{k k^\prime}$ for $k$ and $k^\prime$ at nodes 1 and 2,
respectively.  Note from Eq.\ \ref{delta_e} that a condition for the stability of the superconducting Fermi liquid is that 
$[1 + F^{\mathcal{A}}_{\Gamma}(T)] > 0$.  Thus there is the possibility that the superconducting Fermi liquid will become unstable as the temperature is raised; this is perhaps of interest because of the non Fermi-liquid behavior of the normal state of the high T$_c$ superconductors.

It is also useful to use Eq.\ \ref{S} to obtain an idea of how the 
${\mathcal{S}}$ymmetrical external 
fields are renormalized by Fermi-liquid interactions.
First assume that there are no ${\mathcal{S}}$ymmetrical external
fields other than temperature, i.e. $h_k^{\mathcal{S}} = 0$ 
(as is the case for the external magnetic fields of most interest
in this article, which are purely ${\mathcal{A}}$ntisymmetrical). 
Then the only term driving a nonzero 
contribution to $\delta \varepsilon_k^{\mathcal{S}}$ is the term
on the right hand side 
proportional to $f(E_{k^\prime}^{(0)})$.
This term is proportional to
$T^3$, thus giving a $\delta \varepsilon_k^{\mathcal{S}} 
\propto T^3$, and will not be
important in contributing to the properties of interest at 
the temperatures satisfying $k_B T \ll \Delta_0$ ($\Delta_0$ 
is the maximum gap).  Thus the temperature-dependent contribution to $\delta \varepsilon_k^{\mathcal{S}}$ will be neglected by putting $f(E_{k^\prime}^{(0)})$
on the right hand side of Eq.\ \ref{A} equal to zero. 
In so far as the contributions to $\delta \varepsilon_k^{\mathcal{S}}$ 
driven by $h_k^{\mathcal{S}}$ are concerned, it is clear from Eq.\ \ref{S}
that a knowledge of the Fermi-liquid interaction on the entire Fermi surface
is required to calculate them, and that $\delta \varepsilon^{\mathcal{S}}_k$ must be determined on the entire Fermi surface 
(cf. Ref.\ \protect\onlinecite{mil98}).  To obtain a rough idea of
the nature of the solutions, consider a circular Fermi surface of
radius $k_F$ and look
for a solution of $A_g$ symmetry by considering a Fermi liquid interaction
$f^{(+)}_{k k\prime} = f^{\mathcal{S}}_{A_g}$, independent of $k$ and
$k^\prime$, and a ${\mathcal{S}}$ymmetrical external 
field $h^{\mathcal{S}}_{A_g}$ independent of $k$. The solution, which is
also independent of $k$ on the Fermi surface, gives the renormalized
$h^{\mathcal{S}}_{A_g}$ as
\begin{equation}
	h^{\mathcal{S}}_{A_g} + \delta \varepsilon^{\mathcal{S}}_{A_g} = 
	h^{\mathcal{S}}_{A_g}/[1+ F^{\mathcal{S}}_{A_g}],
	\label{SAg}
\end{equation}
where $F^{\mathcal{S}}_{A_g} = f^{\mathcal{S}}_{A_g} k_F/(\pi \hbar v_F)$.
In contrast to the ${\mathcal{A}}$ntisymmetrical Fermi liquid parameters
$F^{\mathcal{A}}_{\Gamma}(T)$ obtained above, which go to zero linearly with
temperature in the superconducting state in the clean limit (and
hence have a dependence on temperature $T$ explicitly indicated), 
the ${\mathcal{S}}$ymmetrical Fermi liquid parameter
$F^{\mathcal{S}}_{A_g}$ is temperature independent and of approximately
the same magnitude as the corresponding normal state Fermi liquid
parameter.  The same can be seen to be true of the
${\mathcal{S}}$ymmetrical Fermi liquid parameters corresponding to
other irreducible representations of $C_{4v}$.  Note that the ratio
of the ${\mathcal{A}}$ntisymmetrical to the ${\mathcal{S}}$ymmetrical
Fermi-liquid $F$ parameters is 
$F^{\mathcal{A}}/F^{\mathcal{S}} \approx 
(f^{\mathcal{A}}/f^{\mathcal{S}})[k_B T/(\hbar k_F v_2)]$.

\subsection{penetration depth}
As noted above, the presence of a superfluid momentum contributes an
${\mathcal{A}}$ntisymmetrical external field to the Hamiltonian of
Eq.\ \ref{H}.  This external field corresponds to the E irreducible
representation of $C_{4v}$ with the $p_{sx}$ and $p_{sy}$ components of 
${\bf p}_s$ corresponding to the components $Ex$ and $Ey$ of 
Eq.\ \ref{irreps}.  The temperature-dependent quasiparticle contribution
to the current density is thus easily evaluated using
Eq.\ \ref{Jsup} with Eqs.\ \ref{Hdiagonal}, \ref{delta_e} and
\ref{fGamma}. The result is ${\bf J}_{qp} = \eta_{qp} {\bf p}_s$ 
where
\begin{equation}
	\eta_{qp}(T) =  -\frac{2 ln2 e (v^t_F)^2 k_B T}
	{[1 + F^{\mathcal{A}}_E(T)]\pi \hbar^2 \bar{d} v_F v_2}.
	\label{rho}
\end{equation}
Here, $v^t_F$ is the magnitude of the electrical transport velocity at a node, and $\bar{d}$ is the average spacing of the copper-oxide planes.
Note that the Fermi liquid correction does not 
alter the clean-limit linear-in-T contribution to the $\eta_{qp}(T)$,
but rather makes a $T^2$ contribution (using
$(1+F)^{-1} \approx (1 - F + ...)$).  Thus there are no external field renormalization
corrections to the experimentally measured linear in $T$ contribution to
inverse square penetration depth. Hence the low temperature London penetration depth $\lambda$ is given by 

\begin{equation}
	\frac{1}{\lambda_L(T)} = \frac{1}{\lambda^2(0)} 
	- \frac{8 ln2 e^2}{c^2 \hbar^2 \bar{d}} \alpha^2 \frac{v_F}{v_2} k_BT
	+ ...
 	\label{lambda}
\end{equation}
where $\alpha = (v^t_F/v_F)$.

For an isotropic Fermi surface and isotropic Fermi-liquid interactions,
${\bf v}^t_k = {\bf v}_k(1+\frac{1}{2}F_1^s)$, as described in the above discussion of isotropic $s$-wave superconductors (except for the factor 
$\frac{1}{2}$ which arises due to our consideration here of a two-dimensional plane).  By making use of this result, the result of Eq.\ \ref{lambda}, specialized to the case of an isotropic Fermi surface and Fermi-liquid interactions, can be seen to agree with the results of Refs.\ \protect\onlinecite{mil98} and \protect\onlinecite{dur00}. 
A formula for the penetration depth for the anisotropic case has been derived independently in Ref.\ \protect\onlinecite{ran00}.  In their result,
our parameter $v^t_F$ is replaced by an expression depending on $v_F$ and on
the Fermi-liquid interaction paramenters $f^{\sigma \sigma^\prime}_{k k^\prime}$.  The methods of this article can not make a connection between these
two different expressions.

\subsection{Spin susceptibility}
The renormalization of the spin susceptibility due to Fermi-liquid 
interactions can be calculated in a similar way.  The
Zeeman interaction of the spin of an electron with the magnetic field
contributes an ${\mathcal{A}}$ntisymmetric external field of $A_g$
symmetry to the Hamiltonian.  It follows that the magnetic moment
per unit area of a copper oxide plane is
\begin{equation}
	M = -\frac{\mu_B}{L^2} \sum_k \left[f(E_{k,1}) - f(E_{k,-1}) \right]
	= \chi H
	\label{M}
\end{equation}
where 
\begin{equation}
	\chi (T) = \frac{\chi_0 (T)}{1 + F^{\mathcal{A}}_{A_g}(T)},\ \ 
	\chi_0 (T) = \frac{\mu_B^2 ln 2 k_B T}{\pi \hbar^2 v_F v_2}.
	\label{chi}
\end{equation}
Note that here also the low-temperature clean-limit linear in T magnetic
susceptibility is not changed by Fermi-liquid interactions. These
affect only terms of order $T^2$ and higher in the susceptibility.

\subsection{Mixed-state density of states and specific heat}
In the clean limit for a $d$-wave superconductor, the quasiparticle
density of states at zero energy in a magnetic field \protect\cite{vol93} of magnitude $H$ normal to the copper-oxide planes varies as $H^{1/2}$. Electron-electron interactions affect the magnitude of this $H^{1/2}$ contribution to the density of states (and hence to the low-temperature specific heat) and this effect will now be calculated.

It follows from Eqs.\ \ref{Hdiagonal} and \ref{delta_e} that the zero-energy density of states in a
single copper-oxide plane of a $d$-wave superconductor in a magnetic field (characterized by the superfluid momentum ${\bf p}_s$) is
\begin{eqnarray}
 N(0)& = &\left\langle 2\int \frac{d^2 k}{(2\pi)^2} 
	\delta(E_k^{(0)} + ({\bf v}_k^t \cdot {\bf p}_s)/
		[1 + F^{\mathcal{A}}_E(T)] ) \right\rangle
		\nonumber 	\\
	& = & \frac{4 \langle |{\bf v}_F^t \cdot {\bf p}_s| \rangle}
		{\pi\hbar^2 v_F v_2[1 + F^{\mathcal{A}}_E(T)]} 
	\label{N(0)}
\end{eqnarray}
where ${\bf v}_F^t$ is the value of ${\bf v}_k^t$ for ${\bf k}$ at a node.
Here it is assumed that the superconductor is in the mixed state, and the
angular brackets indicate an average of the superfluid momentum ${\bf p}_s$
over a unit cell of the vortex lattice.
This density of states will give rise to a low-temperature specific heat varying linearly with temperature $T$.
As in the evaluation of the low-temperature contribution to the penetration depth above, the quantity $F^{\mathcal{A}}_E(T)$ (which represents the
renormalization of the external field ${\bf v}_k^t \cdot {\bf p}_s$ by the Fermi-liquid interactions), can be neglected as it varies linearly with $T$ and
will contribute only to the next order $T^2$ contribution to the specific heat.
Eq.\ \ref{N(0)} can thus be seen to differ from previous work \protect\cite{vol93,kub98} only by the factor $\alpha = v_F^t/v_F$, which takes into account the different effects of the electron-electron interaction on the electrical transport velocity and the quasiparticle velocity.  (This factor $\alpha$ is the same $\alpha$ that appears in the penetration depth formula, Eq.\ \ref{lambda}, above.)  Thus, the equation for the clean-limit specific heat per unit volume in a magnetic field used in Ref.\ \protect\onlinecite{chi00} (see also Refs.\ \protect\onlinecite{vol93} 
and \protect\onlinecite{kub98}) becomes
\begin{equation}
	\frac{C_{el}}{T} = K\frac{4 k_B^2}{3 \hbar \bar{d}}\frac{\alpha}{v_2}
		\sqrt{\frac{\pi}{\Phi_0}}\sqrt{H}.
	\label{CH}
\end{equation}
The factor K is a  numerical factor of order unity that depends on the way that ${\bf p}_s$  (which is a function of position in the vortex lattice) is averaged over its position coordinate.\protect\cite{vol93,kub98}

In the absence of a magnetic field, there are no Fermi-liquid corrections to the low-temperature specific heat, which is given by the well-known result \protect\cite{chi00} 
\begin{equation}
	C_{el} = 18\zeta (3)k_B^3 T^2/(\pi \bar{d} \hbar^2 v_F v_2).
	\label{C}
\end{equation}	  

\section{Discussion and conclusions}

This article has developed an elementary phenomenological approach to the evaluation of equilibrium Fermi liquid effects in superconductors with anisotropic Fermi surfaces and energy gaps.  The approach is somewhat different from that used in a number of other articles, but gives a formula for the penetration depth that is in agreement with the results of Refs.\ \protect\onlinecite{lar63}, \protect\onlinecite{leg65}, \protect\onlinecite{mil98} and \protect\onlinecite{dur00} for the case of isotropic Fermi surfaces and Fermi-liquid interactions. This article extends these results to systems with anisotropic Fermi surfaces and Fermi-liquid interactions. A classification of the external fields in a superconductor into what have been called above ${\mathcal{S}}$ymmetric and ${\mathcal{A}}$ntisymmetric external fields was introduced.
In $d$-wave superconductors the ${\mathcal{S}}$ymmetric 
external fields (including temperature) exhibit strong Fermi-liquid renormalization effects to which quasiparticles from the entire Fermi surface contribute (cf. Ref.\ \protect\onlinecite{mil98}), 
while the the ${\mathcal{A}}$ntisymmetric external fields (including the magnetic fields that act on both the orbital and spin degrees of freedom of the quasiparticles) 
exhibit relatively weak temperature-dependent renormalizations that arise from the nodal quasiparticles only, and that
can often be neglected.

Eq.\ \ref{lambda} above  for the London penetration depth  has exactly the same form as Eq. 6 of Ref.\ \protect\onlinecite{chi00}, which finds (from a detailed analysis of
a number of experiments) $\alpha^2 = (v_F^t/v_F)^2 =$ 0.43 for Bi$_2$Sr$_2$CaCu$_2$O$_8$
and $\alpha^2 = (v_F^t/v_F)^2 =$ 0.46 for YBa$_2$Cu$_3$O$_{7 - \delta}$. This implies a value of the electrical transport velocity $v^t_F$ at a nodal point
significantly smaller than quasiparticle velocity $v_F$. 

To appreciate the significance of this result concerning the ratio of $v_F^t$ to $v_F$, some idea of the expected relative magnitudes of $v_F^t$ and $v_F$ must be obtained.  Recall that the model studied in this article is that of a single band of electrons interacting with each other and with a periodic potential.  All of the interactions of the electrons in this single band with electrons in other bands or with the ion cores are incorporated into the underlying periodic potential.  The solution of the problem of an electron in this periodic potential in the absence of any electron-electron interactions gives what was called the bare band energy $\varepsilon^b_k$ and the bare band velocity 
${\bf v}^b_k = \hbar \partial \varepsilon_k / \partial {\bf k}$.  At this level of approximation the quasiparticle velocity and the electrical transport velocity are the same, and equal to the bare band velocity.  Now if electron-electron interactions are added in the Hartree-Fock approximation, the band energy, the Fermi surface, and the quasiparticle velocity will all be modified, but the electrical transport velocity remains equal to the bare band velocity.  Furthermore, if the electron-electron interactions are considered in a higher order of approximation in an isotropic model, there will be further changes to the quasiparticle velocity, but the electrical transport velocity will still remain equal to the bare band velocity \protect\cite{bet69}.  Two studies of the effective mass ratio $m^\ast/m$ in an electron-gas model in a high-density approximation have been summarized in Fig. 10 of Ref.\ \protect\onlinecite{sca}. (For an isotropic electon gas model the effective masses are defined by $v_F = p_F/m^\ast$ and
$v_F^t = p_F/m$.)  These studies show that for high densities $m^\ast/m$ is less than unity by up to 5\%, but when the density is decreased the ratio of $m^\ast/m$ becomes greater than unity, by up to 10\% for the range of parameters studied.  Thus, at lower densities, where stronger correlation effects are expected (because the ratio of the potential energy to the kinetic energy in the electron gas becomes larger), the quasiparticle becomes heavier, i.e. tends to move more slowly, as would naively be expected. It is not easy to fit the experimental results on high $T_c$ superconductors quoted above into the framework of these results. The experimental result implies an effective mass ratio $m^\ast/m$ less than unity. Effective mass ratios less than unity (but by a maximum of 5\%) are found in  Ref.\ \protect\onlinecite{sca}, but only in the higher density and relatively less strongly correlated electron gases. This does not seem to fit very well with the idea of electron correlations that are strong enough to lead to an antiferromagnetic Mott insulating state in the high $T_c$ materials for certain doping levels.  It is probably to be expected that high-density electron gas ideas are not particularly useful in understanding the high $T_c$ materials.  

Another potentially relevant study is that of Ref.\ \protect\onlinecite{mon93}, which has calculated the ratio of the quasiparticle effective mass to a bare band mass in a model relevant to high $T_c$ superconductors (see their Eq. (9) and surrounding discussion).  However the band mass there is envisaged as being derived from a energy band structure closely related to that determined by photoemission spectroscopy, and presumably already includes some effects of the electron-electron interactions.  Thus it is not clear how this quantity might be related to the electrical transport velocity which is required for an interpretation of the penetration depth measurements in the formula given above.
It is nevertheless of interest to note in Fig. 5 of Ref.\ \protect\onlinecite{mon93} that electron interaction effects produce a slight band broadening, which goes in the right direction of accounting for the observed value of $\alpha$.

Fermi-liquid corrections to the spin susceptibility, and to the magnetic field induced zero-energy density of states have also been evaluated above for $d$-wave superconductors at temperatures much less that $T_c$.


\section*{acknowledgements}
I would like to thank A. J. Millis for a number  of helpful emails, and
to acknowledge stimulating discussions with A. C. Durst, P. A. Lee,
M. Randeria, and L. Taillefer. The hospitality 
of P. Nozi\`{e}res, and the Theory Group of the
Institut Laue Langevin, where much of this work was done, is much appreciated,  as is 
the support of the Canadian Institute for Advanced Research 
and of the Natural Sciences and Engineering Research Council of Canada.


\begin{references}

\bibitem{har93} W. N. Hardy, D. A. Bonn, D. C. Morgan, R. Liang, and K. Zhang, 	Phys. Rev. Lett. {\bf 70}, 3999 (1993).

\bibitem{pro91} M. Prohammer and J. P. Carbotte, Phys. Rev. {\bf 43},
	5370 (1991).

\bibitem{mar93} J. A. Martindale, S. E. Barrett, K. E. O'Hara, C. P. Slichter, W. C. Lee and D. M. Ginsberg , Phys. Rev. B {\bf 47}, 9155 (1993).

\bibitem{lee93} P. A. Lee, Phys. Rev. Lett. {\bf 71}, 1887 (1993).

\bibitem{gra96} M. J. Graf, S-K. Yip, J. A. Sauls and D. Rainer, Phys. Rev. B {\bf 53}, 15147 (1996).

\bibitem{tai97} L. Taillefer, B. Lussier, R. Gagnon, K. Behnia and H. Aubin, Phys Rev. Lett. {\bf 79}, 483 (1997).

\bibitem{mol97} K. A. Moler, D. L. Sisson, J. S. Urbach, M. R. Beasley, A.
     Kapitulnik, D. J. Baar, R. Liang, and W. N. Hardy  Phys. Rev. B {\bf 55},
	3954 (1997).

\bibitem{wri99} D. A. Wright, J. P. Emerson, B. F. Woodfield, J. E. Gordon, R. A. Fisher and N. E. Phillips,  Phys. Rev. Lett. {\bf 82}, 1550 (1999).

\bibitem{hos99} A. Hosseini, R. Harris, Saeid Kamal, P. Dosanjh, J. Preston, 	Ruixing Liang, W. N. Hardy, and D. A. Bonn, Phys. Rev. B {\bf 60}, 1349, 	(1999).

\bibitem{wal00} M. B. Walker and M. F. Smith, Phys. Rev. B {\bf 61},
	11285, (2000).

\bibitem{chi00} M. Chiao, R. W. Hill, C. Lupien, L. Taillefer, P. Lambert, R. 	Gagnon, P. Fournier, Phys. Rev. B {\bf 62}, 3554-3558 (2000).

\bibitem{kam00} A. Kaminski, J. Mesot, H. Fretwell, J. C. Campuzano, M. R. 	Norman, M. Randeria, H. Ding, T. Sato, T. Takahashi, T. Mochiku, K. 	Kadowaki, and H. Hoechst, Phys. Rev. Lett. {\bf 84}, 1788 (2000).

\bibitem{val99} T. Valla et al., Science {\bf 285}, 2110 (1999).

\bibitem{car99} E. W. Carlson, S. A. Kivelson, V. J. Emery, and E. Manousakis, 	Phys. Rev. Lett. {\bf 83}, 612 (1999).

\bibitem{lan57} L. D. Landau, Sov. Phys. JETP {\bf 3}, 920 (1957).

\bibitem{bet69} O. Betbeder-Matibet and P. Nozi\`{e}res, Ann. 
Phys. (NY) {\bf 51}, 392 (1969).

\bibitem{lar63} A. I. Larkin, Sov. Phys. JETP {\bf 19}, 1478 (1964).

\bibitem{leg65} A. J. Leggett, Phys. Rev. {\bf 140}, A1869 (1965).

\bibitem{gro86} F. Gross, B. S. Chandrasekhar, D. Einzel, K. Andres, P. J. Hirschfeld, H. R. Ott, J. Beuers, Z. Fisk and J. L. Smith, Z. Phys. B {\bf 64}, 175 (1986).

\bibitem{xu95} D. Xu, S. K. Yip, and J. A. Sauls, Phys. rev. B {\bf 51}, 16233 (1995).

\bibitem{mil98} A. J. Millis, S. M. Girvin, L. B. Ioffe and A. I. Larkin, J. 	Phys. Chem Solids {\bf 59}, 1742 (1998).

\bibitem{dur00} A. C. Durst and P. A. Lee, Phys. Rev. B {\bf 62}, 1270 (2000) .

\bibitem{wal00b} M. B. Walker, cond-mat/0007177

\bibitem{ran00} A. Raramekanti and M. Randeria, cond-mat/00011109.

\bibitem{pin89} D. Pines and P. Nozi\`{e}res, 
	{\it Theory of Quantum Liquids - Volume I}, 
	Addison Wesley (1989).

\bibitem{lif} E. M. Lifshitz and L. P. Pitaevskii, {\it Statistical Physics, 	Vol. II}, (Pergamon Press, 1991).

\bibitem{vol93} G. Volovik, JETP Lett. {\bf 58},469 (1993).

\bibitem{kub98} C. K\"{u}bert and P. J. Hirschfeld, Solid State Commum. {\bf 105}, 459 (1998).

\bibitem{sca}	D. J. Scalapino, in {Superconductivity},	
	edited by R. D. Parks (Dekker, 1969), Vol. I, p. 449.

\bibitem{mon93} P. Monthoux and D. Pines, Phys. Rev. B {\bf 47}, 6069 (1993).


\end{references}
\end{document}